\documentclass[12pt,letterpaper]{article}
\usepackage{times}
\usepackage{amssymb}
\usepackage{amsmath}
\usepackage{t1enc}
\usepackage[latin1]{inputenc}
\usepackage[english]{babel}
\usepackage{graphicx}
\renewcommand{\baselinestretch}{1.75}

\begin{document}

\title{Testing the Regular Variation Model for Multivariate Extremes with Flexible Circular and Spherical Distributions}
\author{Fern\'andez-Dur\'an, J.J. and Gregorio-Dom\'inguez, M.M. \\
ITAM\\
E-mail: jfdez@itam.mx}
\renewcommand{\baselinestretch}{1.00}
\date{}
\maketitle

\renewcommand{\baselinestretch}{1.75}

\abstract{
The regular variation model for multivariate extremes decomposes the joint distribution of the extremes in polar coordinates in terms of the angles
and the norm of the random vector as the product of two independent densities: the angular (spectral) measure and the density of the norm.
The support of the
angular measure is the surface of a unit hypersphere and the density of the norm corresponds to a Pareto density. The dependence structure
is determined by the angular measure on the hypersphere, and directions with high probability characterize the dependence structure among the elements of the
random vector of extreme values. Previous applications of the regular variation model have not considered a probabilistic model for the angular density and no statistical
tests were applied. In this paper, circular and spherical distributions based on nonnegative trigonometric sums are considered flexible probabilistic models
for the spectral measure that allows the application of statistical tests to make inferences about the dependence structure among extreme values. The proposed methodology
is applied to real datasets from finance.
}

\textbf{Keywords}: Circular Distribution, Extreme Value Index, Independence Test, Multivariate Circular Distribution, Polar Transformation, Spherical Data

\newpage

\section{Introduction}

For a single random variable, $X$, extreme values are realizations of $X$ that occur with very small probability in the tails of its density function, i.e.,
extreme values are observed with a low frequency but with very large values.
Extreme value theory (EVT) is related to the study of the asymptotic probabilistic characteristics of extreme values. Originally, EVT was the study of the characteristics
of the asymptotic distribution of the maxima from a random sample of independent and identically distributed random variables. Later, EVT also included the asymptotic density function of values that occur over high thresholds.

EVT has important applications in finance and actuarial science and other areas, such as engineering and environmental sciences. In actuarial science, EVT is important to make the valuation of reinsurance contracts and in finance and risk management for the calculation of the expected shortfall and value at risk (VaR) and asset allocation under
conditions of stress and contagion during a financial crisis. In engineering, the typical example is the determination of the height of a dam, and in environmental sciences, there are
important applications in climate change and the monitoring through time of important variables such as temperature.

In this paper, we are interested in the characterization of extreme values of a $d$-dimensional random vector, $\underline{X}$. We consider that an extreme value of the
random vector $\underline{X}$ occurs when the norm of the vector, $||\underline{X}||$, takes very large, extreme values. By considering the norm of the
random vector to determine if an extreme value has occurred, we are allowing the situation in which not necessarily all the $d$ components of the vector $\underline{X}$ take extreme values, and it may be the case that only one component is the one that is extreme. We consider the multivariate regular variation (MRV) model for extremes
that is based on the following result: Let $\underline{X}$ be a $d$-dimensional random vector with distribution $F$. For a random sample of size $n$, as shown by Resnick (1987, page 281), if there exists a real sequence $a_n$ such that
\begin{equation}
nP\{(a_n^{-1} ||\underline{X} ||, ||\underline{X} ||^{-1}\underline{X}) \in (dr,d\underline{s}) \} \rightarrow \alpha r^{-\alpha - 1}dr S(d\underline{s})
\end{equation}
on $(0,\infty) \times \{\underline{x} \in \mathbb{R}^{d} : ||\underline{x}||=1\}$ for a finite measure $S()$ on the $d$-dimensional unit hypersphere, then
$F$ is in the domain of attraction of the extreme value distribution $e^{-\mu([-\underline{\infty},\underline{x}]^c)}$ for $\underline{x}>\underline{0}$, where
\begin{equation}
\mu(\{\underline{x} \in \mathbb{R}^d \setminus {\underline{0}}: ||\underline{x}||>r,||\underline{x}||^{-1}\underline{x} \in A \}) = r^{-\alpha}S(A).
\label{MRVequation}
\end{equation}
Then, for sufficiently large (extreme) values of the norm of the vector of random variables $||\underline{X}||$, the unit norm random vector on the surface of the $d$-dimensional hypersphere, $||\underline{X}||^{-1}\underline{X}$, is independent of the norm random variable $||\underline{X}||=\sqrt{\sum_{k=1}^d X_k^2}$.

Testing the multivariate regular variation (MRV) model for extremes is equivalent to testing for the independence between two sets, $D_1$ and $D_2$, of random variables.
The first set $D_1$ contains the random variables in the unit norm random vector $||\underline{X}||^{-1}\underline{X}$, and the second set $D_2$ only contains the norm
random variable $||\underline{X}||=\sqrt{\sum_{k=1}^d X_k^2}$. Because the random vector in the first set $D_1$ represents a point on the surface of the $d-1$-dimensional hypersphere, $S^{d-1}=\{\underline{x} \in \mathbb{R}^d: ||\underline{X}||=1\}$, it is possible to transform the unit norm random vector
$\underline{X}^*=||\underline{X}||^{-1}\underline{X}$ to spherical coordinates in terms of a set of $d-1$ angles $\Phi_1, \Phi_2, \ldots, \Phi_{d-2}, \Phi_{d-1}$ with
\begin{equation}
\Phi_m = \arccos\left(\frac{X_m^*}{\sqrt{\sum_{k=m}^d X_k^{*2}}} \right)
\label{sphericaltransformation}
\end{equation}
for $m=1,2, \ldots, d-2$ and
\begin{eqnarray*}
\Phi_{d-1}=\arccos\left(\frac{X_{d-1}^*}{\sqrt{X_{d-1}^{*2}+X_{d}^{*2}}} \right) \mbox{ if } X^*_d \ge 0 \\
\Phi_{d-1}=2\pi - \arccos\left(\frac{X_{d-1}^*}{\sqrt{X_{d-1}^{*2}+X_{d}^{*2}}} \right) \mbox{ if } X^*_d < 0 \\
\end{eqnarray*}

The random angles $\Phi_1, \Phi_2, \ldots, \Phi_{d-2}$ take values in the interval $[0,\pi]$ and $\Phi_{d-1}$ in $(0,2\pi]$, and we refer to them as the polar angular random variables. At minimum, the density function of the angle $\Phi_{d-1}$ must have period $2\pi$ and has support on the unit circle. These types of random variables are known as circular (angular) random variables and are studied in the area of circular statistics (see Fisher 1993, Jammalamadaka and SenGupta, 2001, Upton and Fingleton, 1989 and Mardia and Jupp, 2000). We refer to the random variable of the norm of the random vector as the radial component.

The fitting of an MRV model to data consists, first, of the determination of a threshold value for the radial component in the sense that random vectors with a norm larger than the determined threshold value satisfy the MRV condition and, second, of the estimation of the extreme value index (EVI) for the radial component and the estimation of the joint density function of the polar angular random variables (see Mikosch, 2005). The determination of the threshold value is generally done by applying two statistical tests, one for testing for the independence between the radial component and the set of polar angular random variables and the other to test for the regular varying (Pareto) tail of the radial component. In particular, the joint density function of the polar angular random variables, also known as the spectral measure, which corresponds to a density on the surface of a unit hypersphere, contains all the information about the dependence among the extreme values and, in particular, determines directions of preference for the occurrence of extreme values.
There exists many different estimators for the EVI ($1/\alpha$ in Eq. \eqref{MRVequation}) of the radial component. Fedotenkov (2020) reviews more than one hundred methods for the estimation of the EVI. One of the most known is the Hill estimator (Hill, 1975) defined for an ordered $i.i.d.$ sample, $Y_{(1)} \ge Y_{(2)} \ge \ldots \ge Y_{(n)}$ by using the first $r$ largest values as $\hat{\alpha}_{Hill}=\left(\frac{1}{r}\sum_{i=1}^{r} \ln(Y_{(i)}) - \ln(Y_{(r)})\right)^{-1}$. Dekkers, Einmahl and De Haan (1989) developed a moment estimator of the EVI. Alternatively, one can implement the POT (Peaks Over Threshold) method that considers the estimation of the parameters of a Generalized Pareto Distribution (GPD) to the exceedances over the threshold value (see Embrechts, Kl\"{u}ppelberg and Mikosch, 1997).

For the test of the regular varying (Pareto) tail of the radial component, Dietrich, De Haan and H\"{u}sler (2002) and H\"{u}sler and Li (2006) developed methodologies to test if
a distribution belongs to the maximum domain of attraction of an extreme value distribution, known as the extreme value condition, for a dataset without specifying the EVI by considering the Cram\'er-von Mises test statistic for the inverse distribution on the tail.

Recently, a statistical test for MRV was presented in Einmahl, Yang and Zhou (2021), which we refer to as the EYZ test.
The EYZ test consists of two tests: the EYZ test of independence between the radial component and the angular components of the polar representation of the random vector and
the EYZ test for regular varying (Pareto) tails of the radial component. In practice, observed values of the polar random vectors are ordered in decreasing order of the values of their radial component, and both EYZ tests are applied to the first $k$ random polar vectors that correspond to those with the $k$ largest values of the radial component. The two tests (independence and Pareto tail) are applied for various values of $k$ to identify values of $k$ in which the two conditions of the multivariate regular variation model are satisfied.
In the case of a bivariate random vector, the EYZ test of independence consists of testing for the equality of the extreme value index (EVI) on a partition of the interval $[0,2\pi)$, which is the support of the polar angular random variable ($\Phi_1$). The partition is produces approximately the same number of observations in each subinterval of the partition. The number of subintervals of the partition is determined subjectively. Let $\hat{\gamma}_j$ be the estimate of the EVI in the $j$-th subinterval of the partition; then, the EYZ test statistic for independence for $k$ observations, $T_k$, is defined as
\begin{equation}
T_k = \frac{k}{m}\sum_{j=1}^{m}\left(\frac{\hat{\gamma}_j}{\hat{\gamma_{all}}}-1 \right)^2
\end{equation}
where $\hat{\gamma}_{all}$ is the estimate of the EVI for all the observations in the $m$ subintervals of the partition.

The second EYZ test for a regular varying tail of the radial component is the PE test of H\"{u}sler and Li (2006). Einmahl, Yang and Zhou (2021) showed that the two EYZ tests are independent and that a combined p-value can be calculated as $1-(1-\min(p_1,p_2))^2$, where $p_1$ is the p-value of the EYZ independence test and $p_2$ is the p-value of the EYZ regular varying tail test. The main disadvantage of the EYZ test is that the selection of the number of subregions of the partition to test for independence
is specified subjectively and represents a challenge when increasing the dimension of the random vector, and it is necessary to select a partition of subregions on the surface of a high-dimensional hypersphere.

Although the MRV tests are constructed to be used in a random sample (i.i.d.) of random vectors, in practice, the components of the random vector are time series of certain variables and generally present serial correlation, making the occurrence of extreme values occur in nearly consecutive periods. In this sense, Einmahl, Yang and Zhou (2021)
consider that the EZY test is a conservative test meaning that no rejection of the MRV condition in the EZY test implies no rejection when adjusting for serial dependence.
In this same respect, Dacorogna and Samorodnitsky (2001), by analyzing how estimates of the spectral density behave under time aggregation of financial returns, concluded that these estimates are insensitive to serial dependence and that the spectral measure for short and long return intervals behaves very similarly. For the case of constant conditional correlation (CCC) multivariate GARCH models for financial time series, St\u{a}ric\u{a} (1999) developed an estimator of the spectral measure, which depends crucially on a good selection of $k$, by embedding the GARCH model into a stochastic recurrence equation. This paper is divided into five sections, including this introduction. The second section includes the nonnegative trigonometric sum models used to fit the spectral measure and to test the MRV condition. In the third section, we present the proposed statistical test for the MRV condition. The fourth section includes examples of the application of the proposed test to real datasets on financial datasets. Finally, the fifth section includes the conclusions.

\section{Nonnegative Trigonometric Sums Models}

In the simplest case of a bivariate random vector, the spectral measure corresponds to a density on the unit circle and can be estimated by using models for circular densities. In this paper, we consider the use of the flexible family of circular distributions based on nonnegative trigonometric sums (NNTS) (see Fern\'andez-Dur\'an, 2004) to estimate by the maximum likelihood method (see Fern\'andez-Dur\'an and Gregorio-Dom\'inguez, 2010) the spectral measure in the bivariate case. The computational algorithms to fit NNTS models are included in the $CircNNTSR$ package (see Fern\'andez-Dur\'an and Gregorio-Dom\'inguez, 2016) of the $R$ statistical software ($R$ Core Team, 2021). The NNTS family of circular densities allows us to model datasets that present multimodality and skewness and are a very good option to determine the different directions of preference for the occurrence of bivariate extreme values, even in the case of sparse directions (see Meyer and Winterberger, 2021). The NNTS density with support interval $[0,2\pi)$ for a circular (angular) random variable $\Theta$ is defined as
\begin{equation}
f_{\Theta}(\theta) = \frac{1}{2\pi}\left|\left|\sum_{k=0}^M c_k  e^{ik\theta}\right|\right|^2 = \frac{1}{2\pi}\sum_{k=0}^M\sum_{m=0}^M c_k\bar{c}_m e^{i(k-m)\theta}
\label{nntsdensity}
\end{equation}
where $i=\sqrt{-1}$ and $e^{ik\theta}=\cos(k\theta) + i\sin(k\theta)$. The parameters of the model are the vector of complex numbers $\underline{c}=(c_0, c_1, \ldots, c_M)$ with
$c_k= c_{rk} + ic_{ik}$ and $\bar{c}_k= c_{rk} - ic_{ik}$ satisfy the constraint $\sum_{k=0}^M ||c_k||^2 = 1$, and $M$ is the number of terms in the sum defining the NNTS density. The parameter $M$ determines the maximum number of modes of the NNTS density, and in the case of a bivariate MRV model, it is related to the number of different directions of preference for the occurrence of extreme values. Further properties of the NNTS densities can be consulted in Fern\'andez-Dur\'an (2004 and 2007).

In the case of a trivariate random vector, the spectral measure is a density on the surface of the three-dimensional unit hypersphere, and the family of spherical nonnegative trigonometric sums models (SNNTS) can be used (Fern\'andez-Dur\'an and Gregorio-Dom\'inguez, 2014). The SNNTS density for a vector $\underline\Theta=(\Theta_1,\Theta_2)$ with $\Theta_1 \in (0,2\pi]$ and $\Theta_2 \in (0,\pi]$, $f(\theta_1,\theta_2;\underline{c},M_1,M_2)$ is defined as

\begin{eqnarray}
f(\theta_1,\theta_2; \underline{c},M_1,M_2) & = & \frac{\sin(\theta_2)}{4\pi}\left\|\sum_{k_1=0}^{M_1}\sum_{k_2=0}^{M_2}
c_{k_1k_2}e^{i(k_1\theta_1 + k_2\theta_2)}\right\|^2 \\
 & = &\frac{\sin(\theta_2)}{4\pi}\sum_{k_1=0}^{M_1}\sum_{m_1=0}^{M_1}\sum_{k_2=0}^{M_2}\sum_{m_2=0}^{M_2}
c_{k_1k_2}\bar{c}_{m_1m_2}e^{i[(k_1-m_1)\theta_1 + (k_2 -m_2)\theta_2] \nonumber }
\end{eqnarray}

The complex vector of parameters $\underline{c}=(c_{00},c_{01},c_{10},c_{11}, ldots, c_{M_1M_2})$ must satisfy the following constraint:
\begin{equation}
\frac{1}{2}\sum_{k_1=0}^{M_1}\sum_{k_2=0}^{M_2}\sum_{\substack{ m_2=0 \\ | k_2 - m_2 | \ne 1 }}^{M_2} c_{k_1k_2}\bar{c}_{k_1m_2}\left(\frac{1 + \cos((k_2 - m_2)\pi)}{1 - (k_2 - m_2)^2}  \right) = 1.
\label{constraint}
\end{equation}
The total number of free $c$ parameters is $2(M_1 + 1)(M_2 + 1) - 2$, and $c_{00}$ is restricted to be a real number.

The proposed test of the MRV condition requires that pairwise independence among elements of a random vector implies joint (groupwise) independence. This requirement is satisfied for the multivariate nonnegative trigonometric sums (MNNTS) distribution. In this sense, if the vector $(\Phi_1^*,\Phi_2^*, \ldots, \Phi_{d-2}^*, \ldots, \Phi_{d-1}, ||\underline{X}|| (mod 2\pi)))$ can be approximately modeled as a multivariate nonnegative trigonometric sum distribution where $\Phi_k^*$ is the angle obtained from the spherical transformation (see Eq. \eqref{sphericaltransformation}) when considering the
rearranged vector $\underline{X}_{arr}=(X_{k+1}, X_{k+2}, \ldots, X_{d}, X_1, X_2, \ldots, X_{k-1}, X_{k})$ such that $X_k$ is the last element in the vector and the spherical transformation
produces an angle in the interval $(0,2\pi]$ with a periodic circular density for the last element of the vector $\underline{X}_{arr}$. The MNNTS (multivariate NNTS) density on a d-dimensional ($d>2$) hypertorus by Fern\'andez-Dur\'an and Gregorio-Dom\'inguez (2014b) for a vector of angles, $\underline{\Theta}=(\Theta_1,\Theta_2, \ldots, \Theta_d)^\top$, is defined as
\begin{eqnarray}
f_{\underline{\Theta}}(\underline{\theta})& = & \frac{1}{(2\pi)^d}\underline{c}_{12 \cdots d}^H\underline{e}\underline{e}^H\underline{c}_{12 \cdots d} \\
& = & \frac{1}{(2\pi)^d}\sum_{k_1=0}^{M_1}\sum_{k_2=0}^{M_2} \cdots \sum_{k_d=0}^{M_d}\sum_{m_1=0}^{M_1}\sum_{m_2=0}^{M_2} \cdots \sum_{m_d=0}^{M_d} c_{k_1 k_2 \cdots k_d}\bar{c}_{k_1 k_2 \cdots k_d}e^{\sum_{r=1}^d i(k_r-m_r)\theta_r} \nonumber \\
\label{mnntsdensity}
\end{eqnarray}
where $\underline{c}_{12 \cdots d}$ is a $d$-dimensional parameter vector of complex numbers of dimension $2\prod_{r=1}^{d}(M_r+1) - 1$ with subindexes given for all the combinations (Kronecker products) of the $d$ vectors $\underline{M}_r=(0,1, \ldots, M_r)^\top$ for $r=1,2, \ldots, d$, where $M_r$ is the number of terms of the sum in Eq. \eqref{mnntsdensity} for the $r$-th component of the vector $\underline{\Theta}$. The vector $\underline{c}_{12 \cdots d}$ must satisfy
$\underline{c}_{12 \cdots d}^H\underline{c}_{12 \cdots d}=||\underline{c}_{12 \cdots d}||^2=\sum_{k_1=0}^{M_1}\sum_{k_2=0}^{M_2} \cdots \sum_{k_d=0}^{M_d} ||c_{k_1k_2 \cdots k_d}||^2=1$. For identifiability, $c_{00 \cdots 0}$ is a nonnegative real number. The vector $\underline{c}_{12 \cdots d}^H$ is the Hermitian (conjugate and transpose) of vector $\underline{c}_{12 \cdots d}$. The MNNTS family has many desirable properties. The marginal and conditional densities of any order of an MNNTS density are also MNNTS densities, and independence among the elements of the vector $\underline{\Theta}$ is translated into a Kronecker product decomposition in the parameter vector $\underline{c}_{12 \cdots d}$. For example, in the trivariate case $\underline{\Theta}=(\Theta_1, \Theta_2, \Theta_3)^\top$, if $\Theta_1$, $\Theta_2$ and $\Theta_3$ are jointly independent, then $\underline{c}_{123}=\underline{c}_{1} \bigotimes \underline{c}_{2} \bigotimes \underline{c}_{3}$ where $\underline{c}_1$,
$\underline{c}_2$ and $\underline{c}_3$ are the parameter vectors of the NNTS marginal densities of $\Theta_1$, $\Theta_2$ and $\Theta_3$, respectively. Similarly,
if $\Theta_1$ is groupwise independent of $(\Theta_2,\Theta_3)^\top$, then $\underline{c}_{123}=\underline{c}_{1} \bigotimes \underline{c}_{23}$, where $\underline{c}_{23}$ is the parameter vector of the bivariate MNNTS density of $(\Theta_2,\Theta_3)^\top$. These results apply to higher dimensions. Most importantly to this paper,
in the MNNTS family, pairwise independence implies (groupwise) joint independence. For example, in the trivariate MNNTS density, if $\Theta_1$ is pairwise independent of
$\Theta_2$ and $\Theta_3$ ($\underline{c}_{12}=\underline{c}_{1} \bigotimes \underline{c}_{2}$ and $\underline{c}_{13}=\underline{c}_{1} \bigotimes \underline{c}_{3}$); then, $\Theta_1$ is groupwise independent of $(\Theta_2,\Theta_3)^\top$ ($\underline{c}_{123}=\underline{c}_{1} \bigotimes \underline{c}_{23}$). This property is the basis for the development of the test for MRV developed in this paper. The property that pairwise independence implies (groupwise) joint independence is satisfied by other linear models, such as the multivariate normal distribution, elliptical distributions (Fang, 1990 and McNeil et al., 2015) and models constructed from pair copulas as vine copulas (Aas et al., 2009 and Aas, 2016). For alternative methods of estimation of the spectral measure, one can consider the kernel density estimator of Hall, Watson and Cabrera (1987) or the spectral clustering based on a random $k$-nearest neighbors random graph of Avella-Medina, Davis and Samorodnitsky (2023).

\section{Statistical Test for Regular Variation Model for Multivariate Extremes}

To apply the tests, the data are ordered in accordance with the values of $||\underline{X}||$, and the test of independence and the test of Pareto tail are applied to the $k$ largest values of $||\underline{X}||$ for a large set of consecutive largest values of $k$. The first part of the proposed test consists of testing for the Pareto tail of the radial component. We used the Anderson-Darling test for the generalized Pareto distribution to determine the values of $k$ such that the Pareto tail condition is satisfied. From this set of values of $k$, a $k_u$ value is selected that determines the threshold value $u$ for which observed vectors such that $||\underline{X}||>u$ satisfy the Pareto condition. In this work, the Hill method is used to select $k_u$.
The second part consists of testing for independence between the random radial component, $||\underline{X}||$, and the random spectral angles, $\Phi_1, \Phi_2, \ldots, \Phi_{d-1}$. Testing for independence between a random variable and a random vector is a particular case of testing for groupwise independence. We assume that the vector $(\Phi_1^*,\Phi_2^*, \ldots, \Phi_{d-2}^*, \ldots, \Phi_{d-1}, ||\underline{X}|| (mod 2\pi)))$ defined above is MNNTS distributed, and then, pairwise independence implies (groupwise) joint independence or it follows a distribution that satisfies the requirement of pairwise independence implying joint (groupwise) independence. Fern\'{a}ndez-Dur\'{a}n and Gregorio-Dom\'{i}nguez (2023) developed pairwise independence tests by using the angular probability integral transforms (APITs) of random variables. These tests showed higher power than empirical copula independence tests (Genest and R\'{e}millard, 2004) when at least one of the variables involved in the test of independence is a circular random variable. Basically, the independence test of Fern\'{a}ndez-Dur\'{a}n and Gregorio-Dom\'{i}nguez (2023) consists of testing for the circular uniformity of the sum (difference) of the angular probability integral transforms of the random variables. For the case of the test of the regular variation multivariate extremes model, this is equivalent to a groupwise test of independence with two sets of random variables, the first one with $d-1$ random variables, $D_1=\{\Phi_1, \ldots, \Phi_{d-1}\}$, and the second one with only one random variable, $D_2=\{||\underline{X}||\}$. Our test of independence consists of testing for pairwise independence for the pairs
$\{APIT(\Phi_k), APIT(||\underline{X}||)\}$ for $k=1, \ldots, d-1$, where $APIT(Y)=2\pi\hat{F}_Y$ is the angular probability integral transform with $\hat{F}_Y$ as the empirical distribution function of the random variable $Y$. The random variables $APIT(\Phi_k)$ for $k=1, \ldots, d-1$ and $APIT(||\underline{X}|| \}$ are all uniformly distributed in the interval $[0,2\pi)$. Fern\'{a}ndez-Dur\'{a}n and Gregorio-Dom\'{i}nguez (2023) used the Rayleigh test of circular uniformity to implement pairwise independence tests based on the sum (difference) of the APITs. The Rayleigh test statistic, $T_{RT}$, for a random sample of angles, $\theta_1, \theta_2, \ldots, \theta_n$, is defined as $T_{RT}=2n\bar{R}^2$ where $\bar{R}$ is the sample mean resultant length. The statistic $T_{RT}$ is asymptotically chi-squared distributed with two degrees of freedom (see Mardia and Jupp, 2000) and implemented in the $R$ package $circular$ (see Agostinelli and Lund, 2022), including small sample corrections in the calculation of the p-value of the Rayleigh test.

The pairwise independence test is derived from the specification of bivariate circular-circular, $(\Theta,\Psi)$ and circular-linear, $(\Theta,Y)$, densities of Johnson and Wehrly (1977) and Wehrly and Johnson(1980), defined as
\begin{equation}
f_{\Theta,\Psi}(\theta,\psi)=g(2\pi F_{\Theta}(\theta) \pm 2\pi F_{\Psi}(\psi))f_{\Theta}(\theta)f_{\Psi}(\psi)=
g(APIT(\theta) \pm APIT(\psi))f_{\Theta}(\theta)f_{\Psi}(\psi)
\label{circularcircular}
\end{equation}
for a pair of circular random variables with $F$ being distribution functions and $f$ being density functions and
\begin{equation}
f_{\Theta,Y}(\theta,y)=g(2\pi F_{\Theta}(\theta) \pm 2\pi F_{Y}(y))f_{\Theta}(\theta)f_{Y}(y)=
g(APIT(\theta) \pm APIT(y))f_{\Theta}(\theta)f_{Y}(y)
\label{circularlinear}
\end{equation}
for circular and linear random variables. The function $g$ is the density function of a circular random variable and corresponds to a copula density (see Nelsen, 1999 and Sklar, 1959).
By considering Eqs. (\ref{circularcircular}) and (\ref{circularlinear}) and the result in circular statistics that the sum of two independent circular uniformly distributed random variables is circular uniformly distributed, the pairwise independence test consists of testing for the uniformity of the sum (difference) of the APITs, modulus $2\pi$, or equivalently, testing for $g$ being a circular uniform density in the interval $[0,2\pi)$, $g(\theta)=\frac{1}{2\pi}$. This result is applied to all the positive (sum of APITs) and negative (difference of APITs) pairwise tests of the radial component with each of the angles defining the spectral measure. To avoid problems with multiple testing of the pairwise tests, the final p-value for the independence between the radial component and spherical angles is obtained by applying the Bonferroni correction (see Goeman and Solari, 2014 and Cinar and Viechtbauer, 2022). For example, in a trivariate case with two spectral angles,
$\Phi_1$ and $\Phi_2$, and the radial component, $||\underline{X}||$, there are four possible tests of pairwise independence, two pairwise tests considering the sum of the APITs
($(2 \pi F_{\Phi_1} + 2 \pi F_{||\underline{X}||})(mod 2\pi)$ and $(2 \pi F_{\Phi_2} + 2 \pi F_{||\underline{X}||})(mod 2\pi)$) and two pairwise tests considering the difference in the APITs ($(2 \pi F_{\Phi_1} - 2 \pi F_{||\underline{X}||})(mod 2\pi)$ and $(2 \pi F_{\Phi_2} - 2 \pi F_{||\underline{X}||})(mod 2\pi)$) for the radial component and each of the spectral angles. Then, four pairwise test p-values are generated, and the Bonferroni correction consists of multiplying each of the p-values by four and taking the minimum as the final p-value of the test of independence. The Bonferroni correction is applied because it allows for dependence among the p-vales and controls the familywise error (FWER), which is the probability that the testing procedure makes even a single erroneous rejection of the null hypothesis (false discovery). Third, all the p-values of the independence test, sums and differences, and the test of Pareto tails of the radial component are combined by applying the Bonferroni correction to obtain the p-value corresponding to each value of $k$ for the joint MRV test. The final value of $k$, $k_{MRV}=k^*$, can be selected as one of the values of $k$ such that the combined joint MRV test is not rejected at a given significance level. For this value $k^*$ and by using the corresponding observed vectors, the final estimates of the EVI index and the spectral measure by fitting NNTS densities are obtained.


Table \ref{Tablesimulation} includes the results of a simulation to study the power of the proposed MRV test. Following Einmahl, Yang and Zhou (2021), we simulated samples for the vector $\underline{X}=(X_1,X_2,X_3)^\top$ from three known multivariate distributions: the trivariate t-Student distribution with zero mean vector and a specified correlation matrix with ones on the diagonal and values $Corr(X_1,X_2)=Corr(X_2,X_3)=s$, and $Corr(X_1,X_3)=0$ and degrees of freedom $\nu=\mbox{0.5 and 2}$, a trivariate Cauchy distribution considering the same correlation matrices as for the t-Student case and zero location vector and a trivariate independent Pareto distribution each of the three components of the vector with the same parameter $\beta=\mbox{0.5, 1 and 2}$. Multivariate t-Student distributions were simulated using the $R$ package $mvtnorm$ (see Genz and Bretz, 2009), and multivariate Cauchy distributions were simulated using the $R$ package $LaplacesDemon$ (see Statisticat, LLC., 2021).

We simulated 1000 samples each of size 1000 and considered the following values of $k=\mbox{250, 300, 350, 400, 450 and 500}$. It is known that the multivariate Student's t test and Cauchy distributions satisfy the MRV condition and that the independent Paretos do not satisfy the MRV condition. Then, the observed power for the multivariate Student's t test and Cauchy distributions in Table \ref{Tablesimulation} corresponds to the expected significance level, that is, to approximately 1000 times the significance level. In contrast, as expected for the independent Pareto case, the MRV null hypothesis was rejected in all 1000 simulated samples. Similar results were obtained for the bivariate equivalent-related Student's t test and Cauchy distributions and the bivariate independent Pareto cases.

\section{Application: Financial Datasets}

For comparison with the multivariate regular variation EYZ test of Einmahl, Yang and Zhou (2021), we apply the proposed multivariate regular
variation (MRV) test on two financial datasets. The first one is on the daily logarithmic returns of the exchange rates of the British pound GPDUSD and the Japanese yen JPYUSD with respect to the USD dollar from January 4, 1999, to July 31, 2009, with 2663 observations. The second is on the daily logarithmic returns of the S\&P500, FTSE and Nikkei indexes from July 2nd, 2001, to June 29th, 2007, with 1399 observations. These two datasets were previously used in the estimation of extreme risk regions by Cai, Einmahl and De Haan (2011) and He and Einmahl (2017), respectively. Both datasets are claimed to satisfy the MRV condition, which is also confirmed by Einmahl, Yang and Zhou (2021) when applying the EYZ test.
Table \ref{Tableestimates} contains the results for the proposed test of the MRV condition for values of $k$ such that $30 \le k \le 300$. The first column includes the names of the series included in the observed random vectors. Columns 2 to 4 present the results of the test for Pareto tails of the radial component at a 5\% significance level, including the threshold value, $u$, the corresponding value $k_u$ and the EVI estimate obtained by using the Hill method as implemented in the $R$ package $evmix$ (see Hu and Scarrott, 2018). Columns 5 to 7 include the selected values of $k_{MRV}=k^*$ among the values that do not reject the null MRV hypothesis at a 5\% significance level used to estimate the NNTS spectral densities in Figs. \ref{bivariatesandp500ftsenikkeiplot}, \ref{bivariatesandp500ftsenikkeiplotB}, \ref{polarplot} and \ref{figuresphere}. This value $k^*$ was selected among the values of $k$ for which the proposed test does not reject the MRV null hypothesis for a significance level of 5\%. The last three columns of Table \ref{Tableestimates} show the value of M ($\underline{M}$) for the best AIC and BIC (M)NNTS fitted models of the spectral density and the final EVI estimate of the radial component at $k_{MRV}=k^*$. The results of the tests are included in Tables \ref{Tableestimates2bonferroni} and \ref{Tableestimates}.

\subsection{British pound GBPUSD and Japanese yen JPYUSD exchange rates}

Einmahl, Yang and Zhou (2021) used four blocks to test the MRV null hypothesis and concluded that the null hypothesis is not rejected for a wide range of values of $k$ less than 200 at a significance level of 5\%. For values of $k$ between 120 and 200, the estimated EVI of the radial component is approximately 0.28.
The top left plot in Fig. \ref{bivariatesandp500ftsenikkeiplot} presents the dispersion plot of the daily logarithmic returns of the JPYUSD, rJPYUSD, in the horizontal axis and GBPUSD, rGBPUSD, in the vertical axis for observed vectors with norms above the Hill Pareto tail threshold of 0.014. The top second plot presents the plot of the p-values against $k$ for the independence test of the radial component and the angle defining the spectral measure. The three horizontal lines represent the values of 10\%, 5\% and 1\%. The p-values are above 5\% (0.05) for values of $k$ between 42 and 273. The top third plot shows the values of the Hill estimates of the EVI, confirming a value of approximately 0.29 for $k$ values of approximately 200, which is close to the value of 0.28 reported by Einmahl, Yang and Zhou (2021). The selected value of $k^*$ among the values of $k$ that do not reject the null MRV hypothesis at a 5\% significance level was equal to $k_{MRV}=k^*=200$.
Finally, the top last plot shows the histogram of the spectral angle and the best BIC and AIC NNTS estimated density for $k_{MRV}=k^*=200$. The estimated NNTS spectral density shows two main dependence directions with modes slightly above 0 and $\pi$ radians, indicating that when extreme values occur, that is, an extreme value of the norm of the vector of logarithmic returns (rJPYUSD, rGBPUSD), it most frequently occurs for a large positive value of the logarithmic return rJPYUSD and not an extreme value for the logarithmic return of the rGBPUSD and a large negative value of the logarithmic return rJPYUSD and not an extreme value for the logarithmic return rGBPUSD.

\subsection{S\&P500, FTSE and Nikkei financial indexes}

First, we performed bivariate analyses for the three pairs of logarithmic returns (rSandP500, rFTSE), (rSandP500, rNikkei) and (rFTSE, rNikkei) and then performed trivariate analysis of the triplet of logarithmic returns (rSandP500, rFTSE, rNikkei). For comparison purposes, the value of $k_{MRV}=k^*$ was selected to be equal to 140 for the three bivariate cases and the trivariate case.

\subsubsection{Bivariate Analyses}

\begin{figure}[h]%
\centering
\includegraphics[scale=.8, bb= 54 144 558 648]{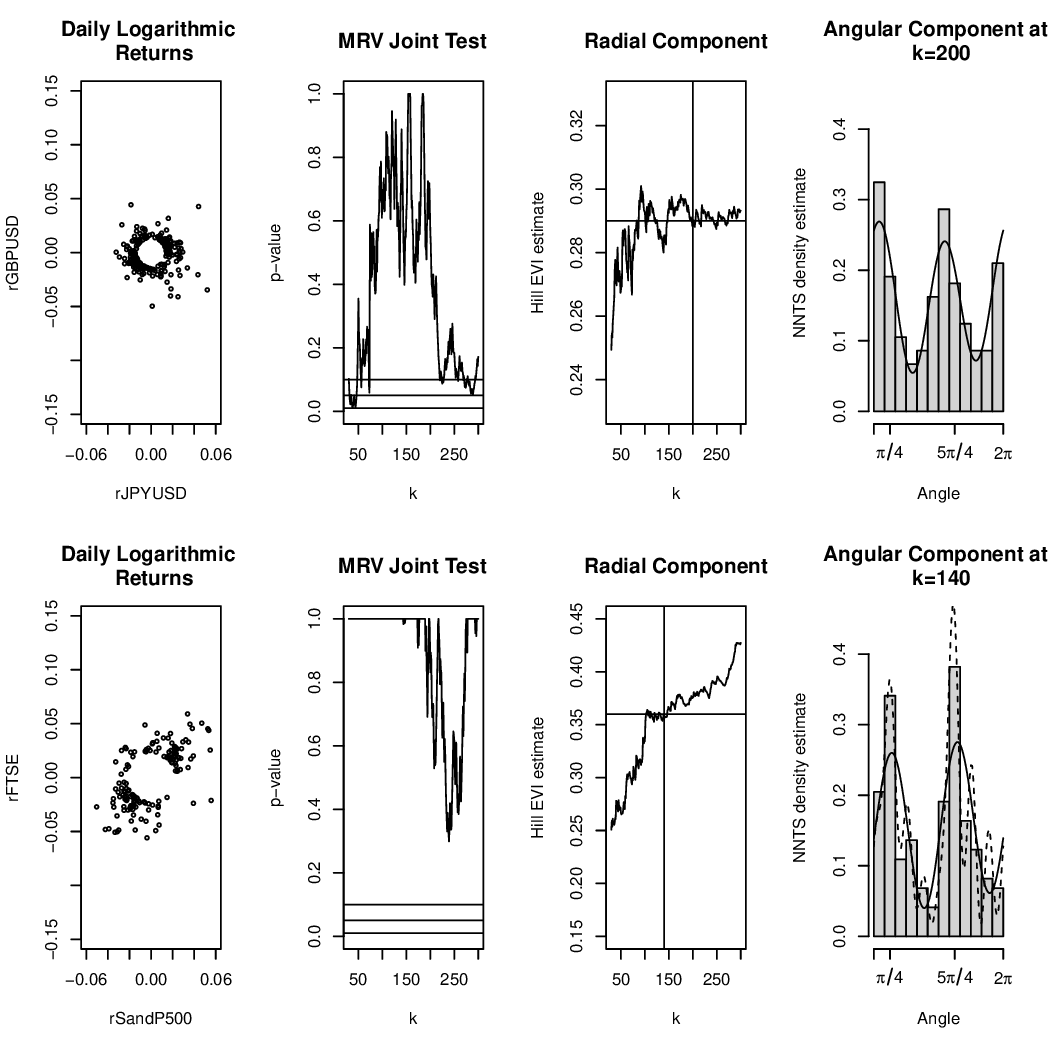}
\caption{Dispersion plot of the observed values above the Pareto tail threshold value in column 2 of Table \ref{Tableestimates} (first column). Plot of the p-values of the independence test between the radial component and the spectral angles against $k$ (second column). Plot of the Hill EVI estimates against $k$ (third column). Best BIC (solid line) and AIC (dotted line) NNTS density estimate of the spectral density $k_{MRV}=k^*$ with the histogram of the spectral angle. The first row includes the plots of the bivariate analysis for (rJPYUSD, rGBPUSD) and the second row for (rSandP500, rFTSE).}
\label{bivariatesandp500ftsenikkeiplot}
\end{figure}

\begin{figure}[h]%
\centering
\includegraphics[scale=.8, bb= 54 144 558 648]{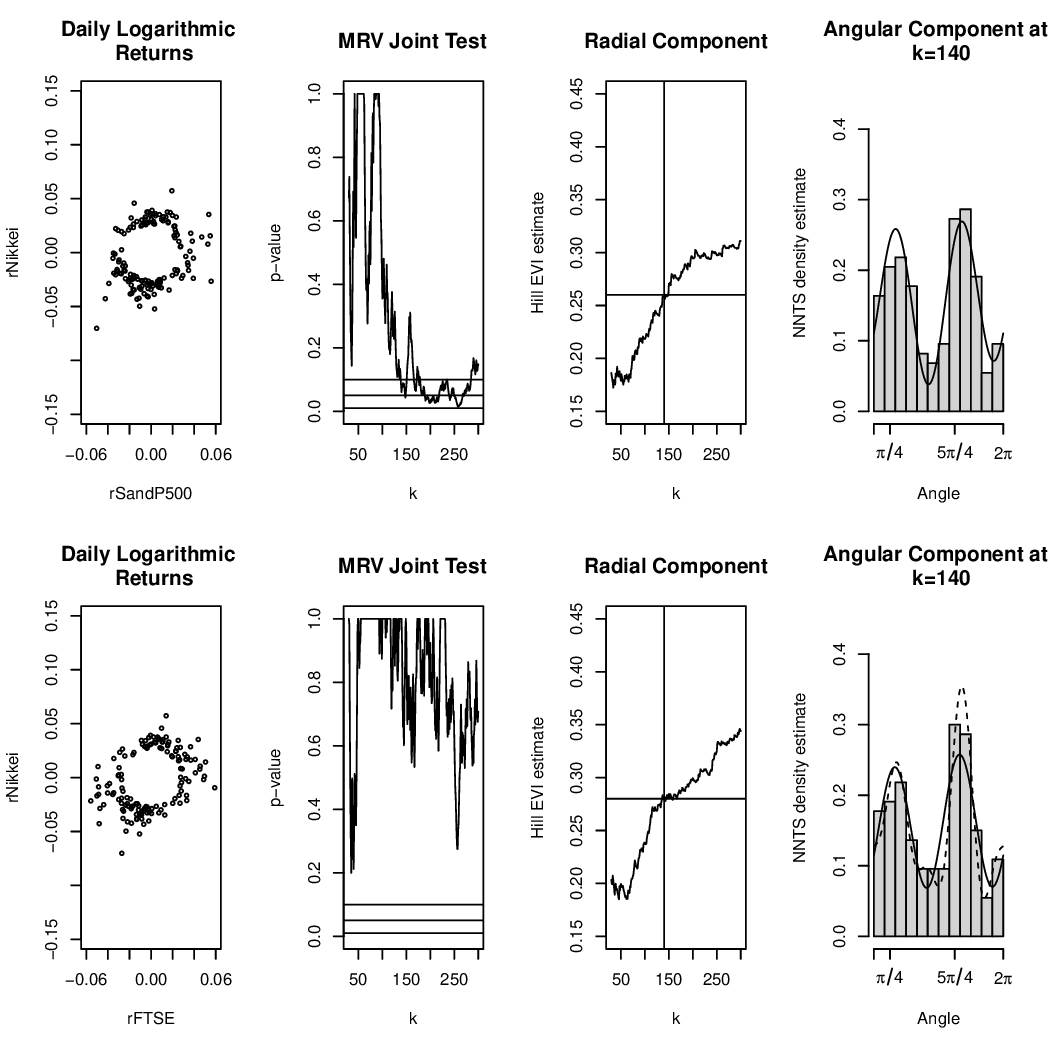}
\caption{Dispersion plot of the observed values above the Hill Pareto tail threshold value in column 2 of Table \ref{Tableestimates} (first column). Plot of the p-values of the independence test between the radial component and the spectral angles against $k$ (second column). Plot of the Hill EVI estimates against $k$ (third column). Best BIC (solid line) and AIC (dotted line) NNTS density estimate of the spectral density at $k_{MRV}=k^*$ with the histogram of the spectral angle. The first row includes the plots of the bivariate analysis for (rSandP500, rNikkei) and the second row for (rFTSE, rNikkei).}
\label{bivariatesandp500ftsenikkeiplotB}
\end{figure}

\begin{figure}[h]%
\centering
\includegraphics[scale=.8, bb= 54 144 558 648]{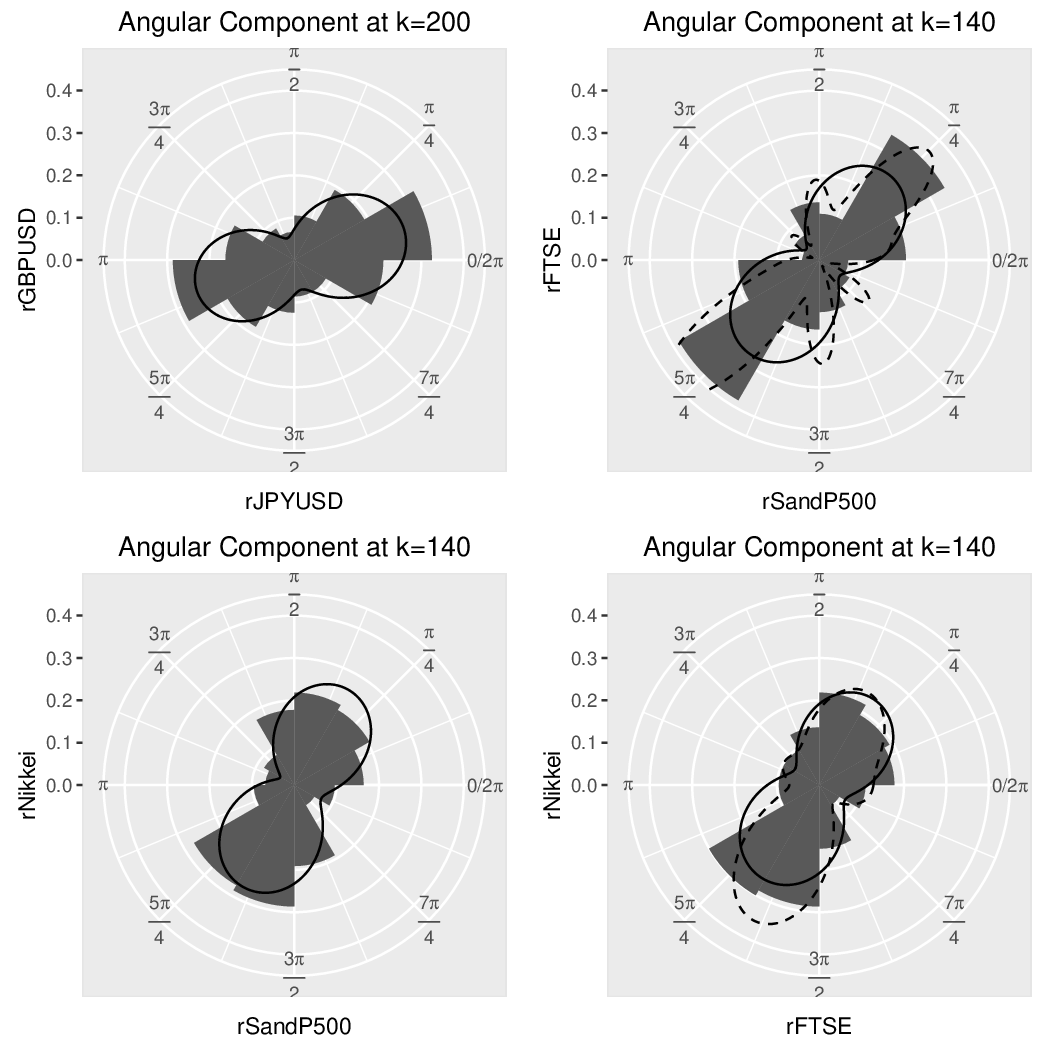}
\caption{Circular histograms and best BIC (solid line) and AIC (dashed line) fitted NNTS densities for the observations satisfying the MRV condition for the pairs
(rJPYUSD, rGBPUSD) top left, (rSandP500, rFTSE) top right, (rSandP500, rNikkei) bottom left and (rFTSE, rNikkei) bottom right. These circular histograms and densities correspond to those in the fourth column of Figs. \ref{bivariatesandp500ftsenikkeiplot} and \ref{bivariatesandp500ftsenikkeiplotB}.}
\label{polarplot}
\end{figure}

The first row of Fig. \ref{bivariatesandp500ftsenikkeiplot} shows the plots for the bivariate analysis of (rSandP500, rFTSE), and Fig. \ref{bivariatesandp500ftsenikkeiplotB} shows the results for the pairs (rSandP500, rNikkei) and (rFTSE, rNikkei).
The first plot of the second row of Fig. \ref{bivariatesandp500ftsenikkeiplot} includes the dispersion plot for the pair (rSandP500, rFTSE) for observed vectors with norms above the Hill Pareto threshold $k_u=\mbox{0.023}$. The second plot shows the independence test plot with p-values that are above 5\% for $k$ values between 30 and 300. The EVI estimates plot is stable at a value of approximately 0.36 for $k$ values between 100 and 150, and the best BIC NNTS estimate of the spectral density for $k_{MRV}=k^*=140$ shows two main dependence directions for modes with similar frequencies at $\frac{\pi}{4}$ and $\frac{5\pi}{4}$, indicating that extreme values occur for simultaneous large positive or negative values of rSandP500 and rFTSE. The Hill EVI estimate for $k^*=140$ is equal to 0.3692.
For the first row of Fig. \ref{bivariatesandp500ftsenikkeiplotB} with the plots for the pair (rSandP500, rNikkei), the independence test p-values are above 5\% for values of $k$ between 30 and 139, and the Hill EVI estimates are stable for values of $k$ approximately 140 for an estimate of 0.28 and the dependence directions with modes at $\frac{\pi}{4}$ and $\frac{5\pi}{4}$ but with different frequencies with larger frequency at $\frac{5\pi}{4}$ than $\frac{\pi}{4}$. The Hill EVI estimate for $k^*=140$ is equal to 0.2899.
The second row of Fig. \ref{bivariatesandp500ftsenikkeiplotB} includes the results for the pair (rFTSE, rNikkei). In the independence test plot, the p-values are above 5\% for values of $k$ between 30 and 300, the EVI estimate is stable at 0.29 for values of $k$ of approximately 145, and the best BIC NNTS spectral density estimate with $k_{MRV}=140$ is very similar to the previous pair (rSandP500, rNikkei). The Hill EVI estimate for $k^*=140$ is equal to 0.2927. Figure \ref{polarplot} shows the circular histograms and fitted AIC and BIC NNTS densities for the considered bivariate analyses.

\subsubsection{Trivariate Analysis}

\begin{figure}[h]%
\centering
\includegraphics[scale=.6, bb= 54 144 558 648]{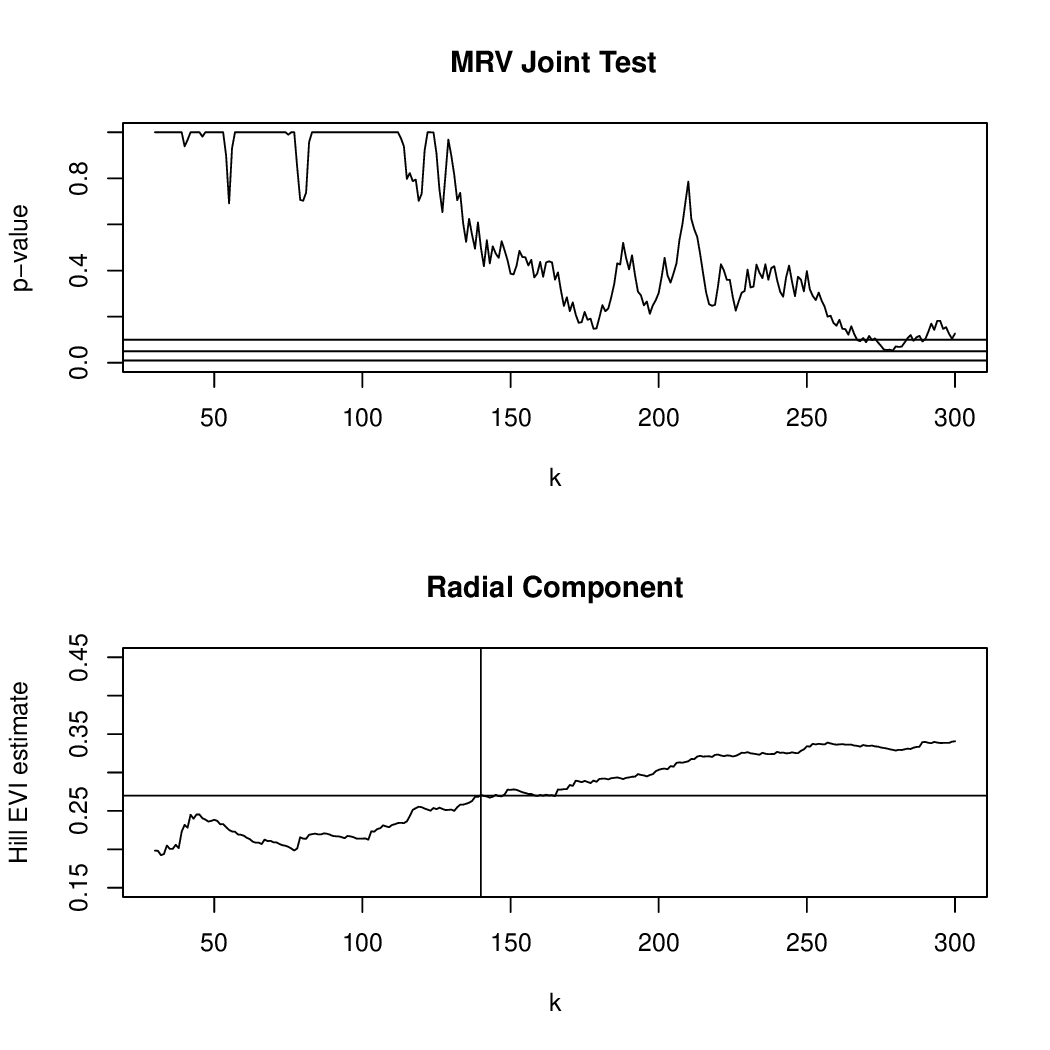}
\caption{Plot of the p-values of the independence test between the radial component and the two spectral angles against $k$ and plot of the Hill estimates of the EVI against $k$ for the trivariate vector (rSandP500, rFTSE, rNikkei).}
\label{trivariatesandp500ftsenikkeiplot}
\end{figure}

\begin{figure}[h]%
\centering
\begin{tabular}{ll}
\scalebox{0.55}{
\includegraphics[0,0][420,315]{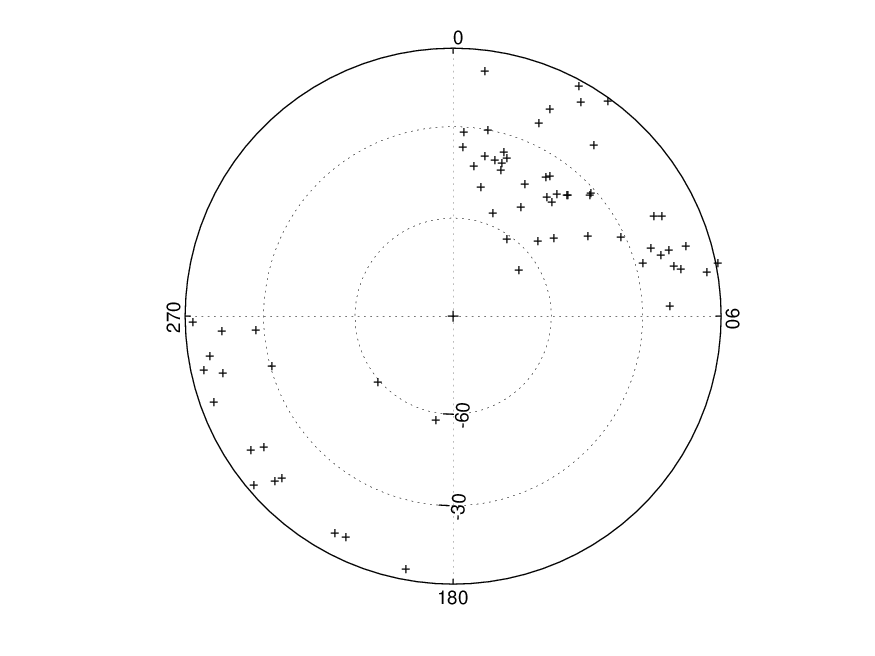}}
&
\scalebox{0.55}{
\includegraphics[0,0][420,315]{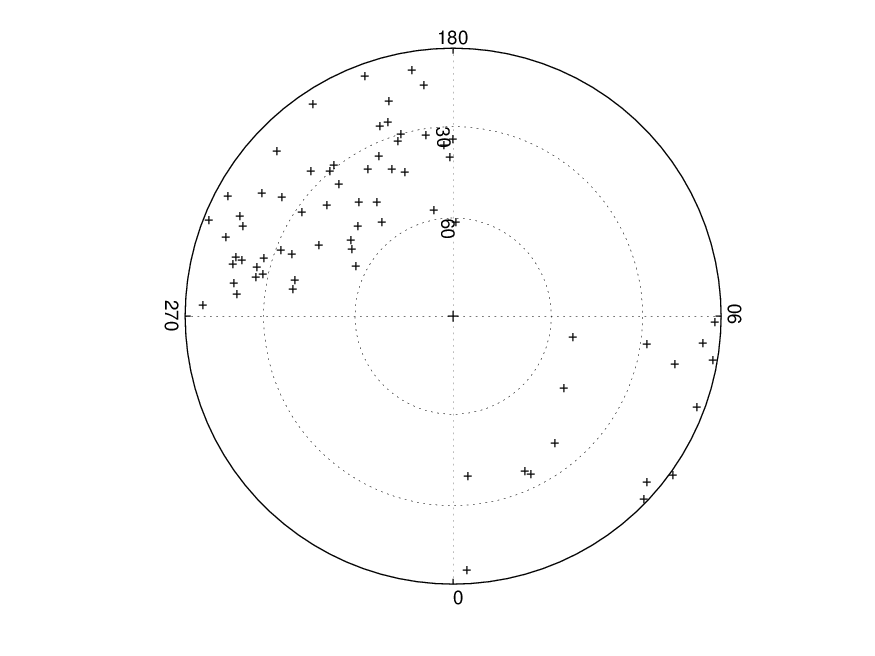}}
\\
\scalebox{0.55}{
\includegraphics[0,0][420,315]{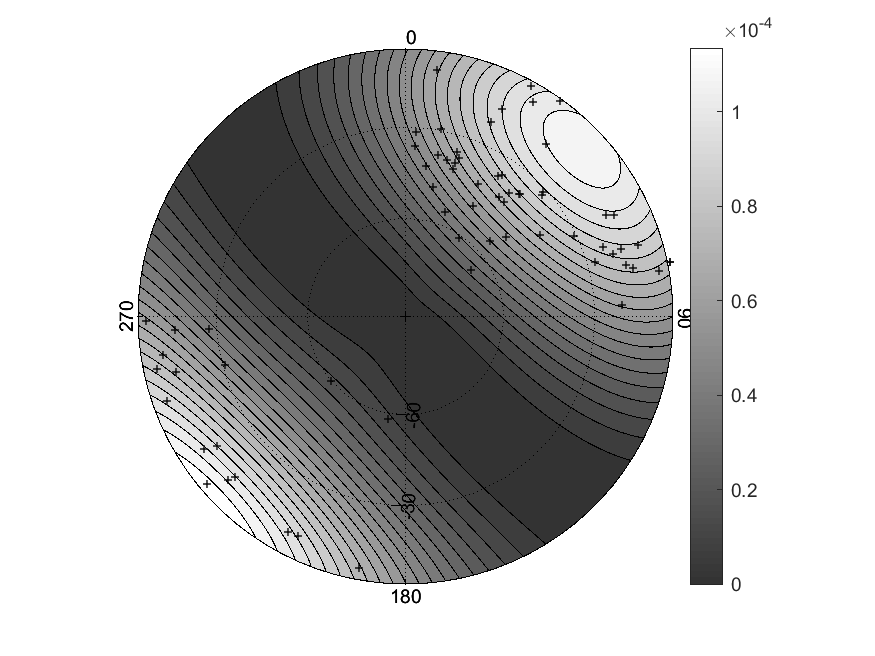}}
&
\scalebox{0.55}{
\includegraphics[0,0][420,315]{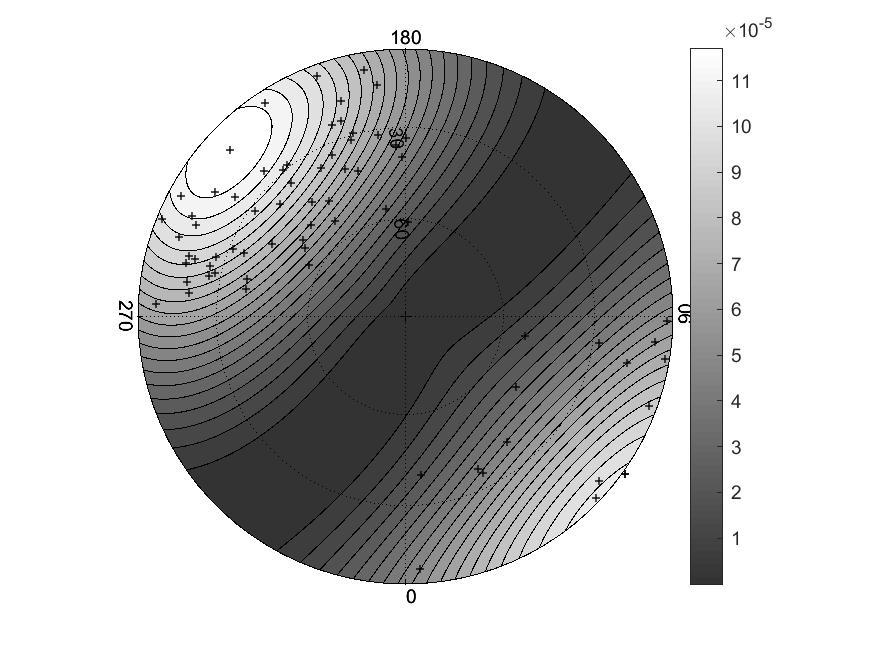}}
\end{tabular}
\caption{SandP500, FTSE and Nikkei logarithmic returns above the Hill Pareto threshold. Best BIC SNNTS Model Plot: The first row of plots presents the data (+) in azimuthal equal-area projection. The plots on the left (right) are for the azimuthal equal-area projection of the north (south) hemisphere. The second row includes the plots of 20 isodensity level curves for the best BIC fitted SNNTS density for the trivariate vector (rSandP500, rFTSE, rNikkei) for the selected value of $k$ not rejecting the null MRV condition at a 5\% significance level ($k_{MRV}=k^*=140$). These plots were created by using the MATLAB software package $\mbox{m}_{-}\mbox{map}$ (Pawlowicz, 2020).}
\label{figuresphere}
\end{figure}

For the triplet (rSandP500, rFTSE, rNikkei), at a significance level of 5\%, \cite{Einmahl2021} do not reject the MRV null hypothesis for values of $k$ less than 150, and for values of $k$ between 50 and 100, they estimated the EVI of the radial component to be approximately 0.24. Figure \ref{trivariatesandp500ftsenikkeiplot} includes the plots of the trivariate analysis for (rSandP500, rFTSE, rNikkei) with the independence test plot indicating that the Bonferroni adjusted p-values are above 5\% for values of $k$ between 30 and 263 with a Hill EVI estimate of 0.27 for $k_{MRV}=k^*$=140.
Figure \ref{figuresphere} in its first row shows the dispersion plots in an azimuthal equal-area projection for the Northern Hemisphere and Southern Hemisphere in which the longitude, latitude and altitude correspond to rSandP500, rFTSE and rNikkei, respectively. From these plots, the dependence among the three logarithmic returns is clear, presenting the expected structure found in the bivariate analysis of (rSandP500, rFTSE). The second row of Fig. \ref{figuresphere} shows the best BIC-fitted SNNTS density in an azimuthal equal-area projection of the Northern and Southern Hemispheres for $k_{MRV}$=140. The fitted SNNTS density has two main modes corresponding to the main directions of dependence among the elements of the trivariate vector, in which it is clear that the strongest dependence of the extremes is among the pair (rSandP500, rFTSE) and that rNikkei generally does not take extreme values when rSandP500 and rFTSE do.

\section{Conclusions}

Testing the multivariate regular variation condition (MRV) is very important for the application of the theory of multivariate extremes to such diverse areas
such as actuarial science, finance, and environmental metrics, among many others. In this paper, a new test for the MRV condition is developed. For the bivariate and trivariate cases, it is possible to estimate the spectral density by using the NNTS and SNNTS models, respectively. The proposed MRV test was applied to datasets in finance and showed good performance.
The proposed MRV test has the advantage that it can be applied to any dimension of the random vector, and computationally, it is efficiently implemented by using the routines of the $CirNNTSR$ package in $R$. One important restriction of the proposed test is that the joint density of the radial component and spectral angles must satisfy that pairwise independence implies joint (groupwise) independence.

\renewcommand{\baselinestretch}{1}
\newpage
\begin{table}[h]
\centering
\begin{center}
\scalebox{0.7}{
\begin{tabular}{|c|c|c|c|c|c|c|c|c|}
\hline
             &            &          & \multicolumn{6}{|c|}{$k$} \\
Distribution & Parameters & $\alpha$ & 250 & 300 & 350 & 400 & 450 & 500 \\  
\hline
Trivariate t-Student & $s=$0 $\nu=$2    & 10\%  & 99  & 100  & 90  & 98  & 97  & 88  \\
                     &                  &  5\%  & 57  &  45  & 44  & 45  & 48  & 42  \\
                     &                  &  1\%  &  9  &  13  &  9  &  6  &  8  &  6  \\
\hline
                     & $s=$0.3 $\nu=$2  & 10\%  & 108  & 92  & 105  & 103  & 98  & 103  \\
                     &                  &  5\%  &  51  & 44  &  50  &  48  & 47  &  47  \\
                     &                  &  1\%  &  15  &  6  &   8  &   8  &  4  &   7  \\
\hline
                     & $s=$0.7 $\nu=$2  & 10\%  & 97  & 93  & 101  & 100  & 124  & 106  \\
                     &                  &  5\%  & 52  & 47  &  61  &  54  &  69  &  58  \\
                     &                  &  1\%  & 15  &  9  &  12  &  15  &  11  &  11  \\
\hline
                     & $s=$0 $\nu=$0.5  & 10\%  & 91  & 87  &  89 &   99  &  84  &  85  \\
                     &                  &  5\%  & 40  & 41  &  40 &   46  &  39  &  36  \\
                     &                  &  1\%  &  1  &  3  &   4 &    0  &   1  &   1  \\
\hline
                     & $s=$0.3 $\nu=$0.5& 10\%  & 86  & 89  &  91 &   90  &  87  &  80  \\
                     &                  &  5\%  & 33  & 42  &  46 &   38  &  24  &  32  \\
                     &                  &  1\%  &  2  &  3  &   2 &    2  &   0  &   0  \\
\hline
                     & $s=$0.7 $\nu=$0.5& 10\%  & 81  & 94  &  88 &  103  &  92  &  78 \\
                     &                  &  5\%  & 33  & 35  &  35 &   42  &  35  &  32 \\
                     &                  &  1\%  &  1  &  1  &   4 &    6  &   6  &   4 \\
\hline
\hline
Trivariate Cauchy    & $s=$0            & 10\%  &101  & 91  & 110 &  110  &  90  & 111  \\
                     &                  &  5\%  & 52  & 41  &  46 &   56  &  39  &  45  \\
                     &                  &  1\%  &  3  &  7  &   5 &    6  &   6  &   3  \\
\hline
                     & $s=$0.3          & 10\%  &100  &100  &  96 &   95  &  92  & 101  \\
                     &                  &  5\%  & 51  & 42  &  42 &   55  &  46  &  38  \\
                     &                  &  1\%  &  4  &  3  &   7 &    9  &   6  &   2  \\
\hline
                     & $s=$0.7          & 10\%  & 97  &104  & 109 &  105  & 113  & 124  \\
                     &                  &  5\%  & 56  & 50  &  49 &   44  &  52  &  65  \\
                     &                  &  1\%  &  7  & 14  &  13 &   15  &  12  &  10  \\
\hline
\hline
Trivariate Independent Paretos & $\beta_1=\beta_2=\beta_3=$0.5 & \multicolumn{7}{|c|}{1000 all cases}   \\
\hline
                               & $\beta_1=\beta_2=\beta_3=$1   & \multicolumn{7}{|c|}{1000 all cases}   \\
\hline
                               & $\beta_1=\beta_2=\beta_3=$2   & \multicolumn{7}{|c|}{1000 all cases}  \\
\hline
\hline
\end{tabular}}
\end{center}
\caption{Power of the proposed MRV test from 1000 simulated samples of size 1000 from the trivariate t-Student, Cauchy and independent Pareto distributions. The reported power is the number of simulated samples in which the MRV null hypothesis is rejected at the specified significance level of 10\%, 5\% and 1\%. The multivariate Student's t and Cauchy distributions satisfy the MRV condition. The independent Pareto distribution does not satisfy the MRV condition.}
\label{Tablesimulation}
\end{table}

\newpage
\begin{table}[h]
\begin{center}
\scalebox{0.8}{
\begin{tabular}{|c|ccc|ccc|ccc|}
\hline
    & \multicolumn{3}{|c|}{GPD Pareto Anderson-}                        & \multicolumn{3}{|c|}{Bonferroni Corrected}      & \multicolumn{3}{|c|}{Bonferroni Corrected} \\
    & \multicolumn{3}{|c|}{Darling (AD) tail test} & \multicolumn{3}{|c|}{Independence Test} & \multicolumn{3}{|c|}{Joint MRV test} \\
\hline
Series & 1\% & 5\% & 10\% & 1\% & 5\% & 10\% & 1\% & 5\% & 10\% \\
\hline
rJPUSD and rGBPUSD            & 30-300  & 30-300  & 30-300  & 45-300  & 47-273  & 64-218  & 39-300  & 47-300  & 74-219  \\
rSandP500 and rFTSE           & 30-300  & 74-300  & 74-300  & 30-300  & 30-300  & 30-232  & 30-300  & 30-300  & 30-300  \\
rSandP500 and rNikkei         & 30-300  & 35-260  & 45-154  & 30-257  & 30-139  & 37-128  & 30-300  & 30-147  & 30-138  \\
rFTSE and rNikkei             & 30-300  & 65-300  & 65-300  & 30-300  & 30-300  & 30-300  & 30-300  & 30-300  & 30-300  \\
rSandP500, rFTSE and rNikkei  & 78-300  & 121-277 & 160-275 & 30-300  & 30-263  & 30-169  & 30-300  & 30-266  & 30-225  \\
\hline
\end{tabular}}
\end{center}
\caption{Results for the proposed test for the MRV condition. Columns 2 to 4 contains the values of $k$ for with the Pareto tail test based on the Anderson-Darling test for the Generalized Pareto Distribution (GPD) is not rejected for significance levels of 1\%, 5\% and 10\%. Columns 5 to 7 presents the range of values of $k$ for which the Bonferroni corrected independence test is not rejected. At last, columns 8 to 10 include the range of values of $k$ for which the proposed Bonferroni corrected joint (Pareto tail and independence) MRV test is not rejected.}
\label{Tableestimates2bonferroni}
\end{table}

\begin{table}[h]
\begin{center}
\scalebox{0.8}{
\begin{tabular}{|c|ccc|ccc|c|}
\hline
    & \multicolumn{3}{|c|}{Hill Method}          & \multicolumn{3}{|c|}{Values Used in Plots} & Hill Method \\
\hline
Series & Threshold & Pareto &      Hill EVI                 &                   & BIC  & AIC & Hill EVI Estimate \\
       & $u$       & $k_u$  & Estimate at $k_u$             & $k_{MRV}=k^*$     & $M$  & $M$ & at $k_{MRV}=k^*$ \\
\hline
rJPUSD and rGBPUSD            & 0.014  & 245  & 0.2851  & 200  & 2  & 2  & 0.2873  \\
rSandP500 and rFTSE           & 0.023  & 146  & 0.3681  & 140  & 2  & 8  & 0.3692  \\
rSandP500 and rNikkei         & 0.026  & 149  & 0.2839  & 140  & 2  & 2  & 0.2899  \\
rFTSE and rNikkei             & 0.027  & 145  & 0.2908  & 140  & 2  & 4  & 0.2927  \\
rSandP500, rFTSE and rNikkei  & 0.032  & 137  & 0.2979  & 140  & 2  & 2  & 0.2936  \\
\hline
\end{tabular}}
\end{center}
\caption{Results of the Hill method and values used in the plots. Columns 2 to 4 contains the threshold value, $u$, the value of $k$ at the threshold $u$, $k_u$ and, the estimated EVI by using the Hill method. Column 5 shows the selected value of $k$, $k_{MRV}=k^*$ from the set of $k$ values in Table \ref{Tableestimates2bonferroni} at which the proposed Bonferroni corrected test does not reject the null MRV condition at a 5\% significance level. Columns 6 and 7 include the values the value of the NNTS parameter $M$ of the estimated AIC and BIC NNTS density at $k_{MRV}=k^*$. At last, column 8 shows the Hill estimate of the EVI at $k_{MRV}=k^*$.}
\label{Tableestimates}
\end{table}

\newpage

\newpage

\newpage

\newpage

\newpage

\end{document}